# Determination of refractive index of various materials on Brewster angle

E. A. Tikhonov, Institute of Physics, National Academy of Sciences, Kiev-28, prospect Nauki, 46, Ukraine, e-mail: etikh@iop.kiev.ua

**Abstract:** Studied experimentally the origin of the non-zero reflection of p-polarized radiation (TM) of Brewster's angle. The results have shown the residual reflected light in the vicinity of Brewster angle occurs due to inaccessibility 100% polarization degree the incident linearly-polarized radiation and installation of the zero azimuthal angle. These factors create the s-component of the radiation reflected from the examined surface indeed. A smooth change of reflected light polarization in the vicinity of Brewster angle in the sequence p-s-p appears due to the changing power proportion of reflected p-, and s-components but not is the result of the atomically thin transitional layer at the border of the material/environment according to Drude model. Metrological aspects of refractive index measurement by Brewster angle are investigated: due to the above-mentioned factors, as well as due to the contribution of the reflected scattered light caused by on residual roughness of the optical surface. Advantages of Brewster refractometry for any materials and films without restrictions on the topology of samples and their light scattering and absorption are demonstrated.

1. **Introduction**

In works [1,2], made on the frame of Fresnel's theory, measurements of refraction index (RI) at normal and Brewster incidence angles of laser radiation were compared. It is noted that Brewster refractometry differs in advantage owing to the single system error in the absence of restrictions for sample topology and RI magnitude. The similar method of ellipsometry [3] cedes Brewster one because the need to measure the whole three parameters: reflectivity s,- p-polarized light power and the selected angle of their registration.

The error of Brewster angle determination has in turn two different reasons. The first is connected installation the zero angle of reference attached to the sample surface. The task is solved by collimation superposition of direct and return laser beam with typically small, (<1') angular divergence. The second reason is connected the installation of zero azimuthal angle ($\alpha=0$) demanding superposition of polarization and incident planes of radiation on the tested surface. At the same time the degree of polarization (DP) of radiation acts as the contributor to this error: it decreases the requirement of installation of a zero azimuthal angle to a value of DP. Thus, both factors reduce the determination accuracy of Brewster angle on a minimum of residual reflected power.

It is appropriate to notice also that the registered residual radiation and change of its polarization in the narrow vicinity of Brewster angle in sequence "p-s-p" was considered as violation of the classical Fresnel theory. This violation was removed at introduction of a hypothesis of availability and influence of a super thin (<<$\lambda$) boundary layer at any material with the correspondent Drude`s addition l to Fresnel theory [4]. The similar monolayer cannot attribute homogeneous on depth RI transition from one media to another.

Drude's model was tested by Rayleigh on a number of liquids which surfaces were considered more optically homogeneous than when rigid bodies. Experiment with change of polarization at reflection in the small vicinity of Brewster angle was treated within Drude's addition of Fresnel theory [4]:

$$\frac{R_p}{R_s} = i[\frac{\pi}{\lambda}\sqrt{n^2+1}(\gamma_z - \gamma_x)] \qquad (1)$$

In a formula (1) ellipticity of the reflected light is determined by the relation of reflectivity $R_{p,s}$ for emissions with p-, to s-polarization depending on RI=n of material, the wavelength of measurement $\lambda$ and a difference of some combinations RI for given the film. Rayleigh has measured the ellipticity value and found the thickness of a transitional layer l for water equal: $R_p/R_s=42*10^{-5}$, $l=0,00057\lambda$. Thus, in Drude's model ellipticity of radiation and component of s-polarization in the small vicinity of Brewster angle arise at the reflection of the incident p-polarized radiation from the super thin transitional layer which does not influence the value of Brewster angle of the main material.

The results of given work taking into account the fatal contribution of s-polarized radiation due to the degree of polarization of emission less than 100% and due to an azimuthal angle never are zero allow to explain the seeming deviation from Fresnel's theory without appeal to a transitional layer and Drude's model based on it.

In the light of the made remarks, we will estimate influence on observed data of RI the nonzero azimuthal angle and less than 100% DP laser emission used in measurements. Really even CW lasers with Brewster elements inside cavity provide radiation with DP= 100% never.

Normalized power of the reflected light as function of an incidence angle $\varphi$, the azimuthal angle $\alpha$ and RI=n is given within Fresnel's model by next expression:

$$\frac{P(\varphi, \alpha.n)}{P_0} = (\frac{\cos\varphi - n\cos\psi}{\cos\varphi + n\cos\psi})^2 \sin^2\alpha + (\frac{n\cos\varphi - \cos\psi}{n\cos\varphi + \cos\psi})^2 \cos^2\alpha \qquad (2)$$

At values α=90⁰ or 0⁰ reflected the power of linearly polarized light p-, and s-orientations are described by the first or second member (2) respectively. The power of reflection of depolarized light or light of circular polarization forms an equal contribution of both members of expression (2). A condition under which the sum of incidence angles and refraction is $\varphi+\beta=\pi/2$, according to Fresnel (2) (at α=0) leads to reducing the reflected power of p-orientation to zero owing to the transversal structure of an electromagnetic field. By name discoverer of the phenomenon, the angle bears Brewster name.

At α≠0, the first member (2) makes a contribution in reflection for account of s-polarized components from the incident p-polarized radiation even at DP=100%. While reflection on Brewster angle for p-polarization seeks to zero, reflection for s-polarization radiation continuously grows with an incidence angle to 100%, the total reflected power creates the minimum shifted towards smaller angles. The magnitude of this shift is proportional to a contribution from the s-polarized components [1,2]. Similarly, the angular position of a minimum of the reflected power (MRP) depends from DP of p-oriented components of testing beam even in a case α=0.

The way out from the similar difficulty, limiting RI measurement accuracy on an angular MRP of p-polarized radiation, consists in the application of polarization analyzer in front of the reflected power receiver with tuning on the suppression of the arising s-polarized component. The made decision was validated at all next RI measurements of different materials, in particular, materials with the rough tested surface.

Following the standard definition of DP=p, from relation (2) at α≠0 one receives expression for DP of reflected beam:

$$p = \frac{R_s(\varphi)\sin^2\alpha - R_p(\varphi)\cos^2\alpha}{R_s(\varphi)\sin^2\alpha + R_p(\varphi)\cos^2\alpha} \qquad (3)$$

From (3) follows that DP in a reflected beam at α≠0 changes with an incidence angle always ($R_{s,p}$ - Fresnel reflectivity from a formula (1)). At the same time on Brewster angle for the s-polarized radiation, irrespective of the size of an azimuthal angle α, p=1. For s-polarized radiation at α=0, the reflected radiation is absent at all if incident radiation is characterized by infinitely high DP.

Further, results of RI measurement and their analysis for materials of different topology with missing or weak absorption when the real part of RI much and more orders exceeds an attenuation index ($n^2 >> \varkappa^2$) will be provided.

## 2. Determination of Brewster angle at one reflecting surface

Determination of Brewster angle for materials of different topology (volume samples with one optical surface, plates, and films with two optical surfaces) were carried out using CW lasers of different wavelengths at minute divergence and power in level (1÷20) mW. The angular step of sample rotation was set by a step motor and reducer and was equal 0, 024 minutes. Measurement of radiation power (in relative units - volts) was performed by means of a handmade photometer on silicon photodiodes PhD-24 in the short-circuit mode when photocurrent is directly proportional to a luminous flux in limits to 3 orders of values [5]. Magnitudes of an analog signal in volts (0,01÷4,5) were digitized by 10-digital ADT and entered into the computer for processing and graphical representation. The level of dark noise was lower 1mv and partially depended on electronic amplification of the photometer. The receiving photodiode together with a studied sample was established on a goniometer head with the scale of reference 1 angular minute. The goniometer allowed to keep the position of a reflected beam on a photodetector at all angular changes of the incident beam. The reference signal from the similar photodetector was detected and applied for correction of indications from the signal channel if power fluctuations of the measuring laser took place. When changes of reflected signal power reached to 3 orders of values, to reduce a nonlinear response of a photometer to input light power an attenuated light filters were used. Starting values of DP and laser power were measured with dichroic polaroid in both channels by above mentioned dual channel photometer. Change of linear beam polarization state from s-, p- orientations concerning the incident plane was performed by rotation of a diode laser around an optical axis on π/2.

Let's address typical results of measurement of MRP angle for p-polarized radiation for glass (K8) received with the described procedure of registration (fig. 1.)

From the aforesaid it is followed that residual power in MRP depends on emission DP and the different from zero azimuthal angle α. Besides, power in MRP depends also on the optical roughness of the surface causing light scattering with some depolarization [6].

On incidence angles up to the ≈20° decrease of DP is practically absent owing to small distinctions of reflectivity (2) at the change of an azimuthal angle. For example, DP of He-Ne incident radiation (632,8nm) is equal 99,92%, but after reflection by an optically polished surface of the K8 glass with azimuthal angle 70° reduces to 99,62% (distinction of 0,33%). Not attenuated part of depolarized components of light will grow with an incidence angle and will be registered in the vicinity of Brewster angle. The procedure of RI measurement at Brewster angle in MRP of p-polarized radiation on the example of K8 glass and λ=632,8nm is presented on fig.1. Without the mentioned polarization filter (analyzer) on a photodetector, the small difference of the measured value RI from provided by Russian state standard is observed: $\varphi_1$=56,333°, RI=1,5013 and $\varphi_2$=56,5660, n=1,5145 respectively. The difference was eliminated at the proper fitting of the polarization analyzer suppressing s-component of reflected radiation.

Let's address the analysis of residual radiation in the vicinity of Brewster angle, important as for the understanding of the physical phenomenon, so for justification of measurement procedure RI at MRP. Today's understanding of the phenomenon was given in the introduction. It is based on a hypothesis of existence super thin transitional layer on the border with the environment (air). RI of such 2-dimensional film can smoothly change from RI value of air to software of a tested material. In some approach, the state of an intermediate layer can be compared to a bulk material with homogeneously distributed pores inside which diameter are smaller than wavelengths of measuring light.

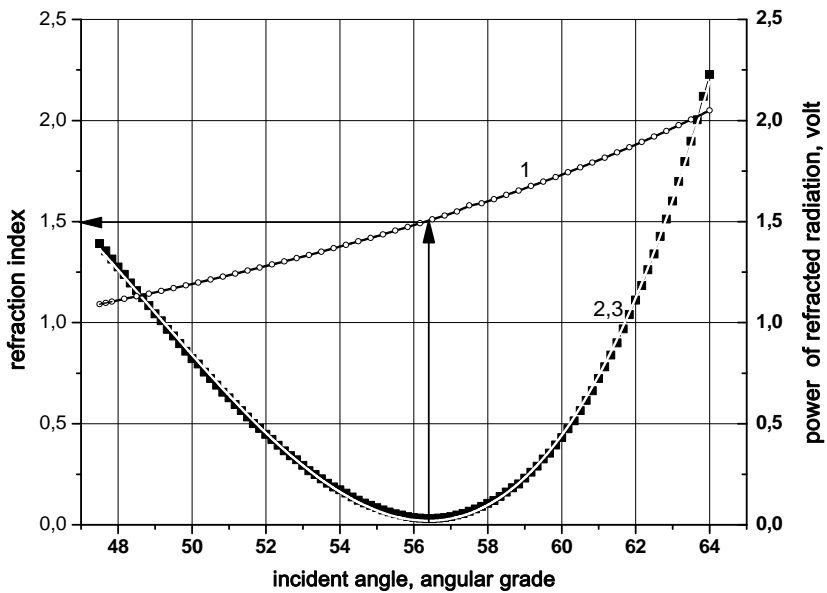

Fig.1. 1- tan($\varphi$)=n, 2-measured values of light power, 3- polynomial of 4-th power (white line across squares, correlation coefficient- 0,9994). Arrows indicate sequence RI determination.

At an increase of pore concentration an efficient RI of similar material changes from RI of continuous material to RI of air [11]. It seems that Brewster light reflection from a similar film will be present at the wide sector of incidence angles as RI similar film will "be smeared" in a wide area between values for boundary homogeneous materials so to register single MRP in given films are not possible.

To study behavior of reflected light of p-polarization and irremovable s-polarized component, their reflection was registered in a wide sector of incidence angles from 10° to 70°. The polarized radiation of the injection diode laser $\lambda=660nm$ with DP= 99,9% and fitting of an azimuthal angle $\varphi \cong 0$ in a plane of incidence on the optical quality surface of glass K8 used. The analyzer on the photodetector was fitted in turn in two orthogonal orientations passing p, - or s-component of polarization of the reflected light. The both dependency (fig.2.) contains the information needed for an understanding of light behavior in Brewster minimum and its influence on RI measurement accuracy. The ratio of the variable reflected radiation power of p, - and s-polarization on the angle shows how high (99,66%) initial of total radiation on small angles DP decreases up to change of a sign in vicinity of Brewster angle. In intersection points 2S- and 1P-dependences of fig.2. in the vicinity of Brewster angle polarization of total radiation disappears at all, within crossing of curves DP becomes elliptic, outside this area DP of total radiation recovers initial value.

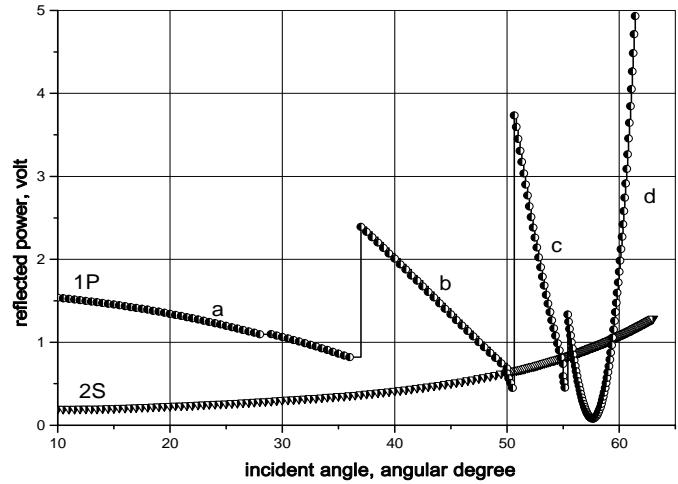

Fig. 2. Angular dependence of the reflected radiation power for K8 glass: the DP after reflection for an angle of 10° - 99,66%, 1p-dependence is power of p-polarized part, 2S-power s-polarized components, plots a, b, c, d on curve 1P correspond to points of increase of the input power: 70,73x24,4x2,9x1=5004,85 times;

The behavior of the total power (1P+2S) considered as p-polarized radiation with DP=100% led to the conclusion that in Brewster's angle vicinity the radiation experiences phase jump $\pi$-radian followed the emergence of ellipticity for reflected nonzero power of p-polarization on Brewster angle. The similar behavior of the radiation with DP=100% and $\alpha=0$ disagrees to the main Fresnel model when polarization jump on $\pi$ on Brewster angle is not followed with the reflected radiation completely.

In our measurements of the reflected p-polarized radiation (a curve 1P, fig. 2.) no additional changes of its power due to polarization jump are revealed. While the total power dependence 1P+ 2S in fig. 2. has all signs of similar behavior, i.e. change of polarization in the narrow vicinity of Brewster angle and appearance of some small reflected power.

For experimental verification of our measurements and, respectively, conclusions, we measured phase jump of the p-polarized radiation with the azimuthal l angle of α≅0 at single TIR on a rectangular prism from K8 glass. In Fresnel's theory the magnitude of phase jump is defined for the case by next expression [4]:

$$\delta_p = 2\arctan\left(\frac{\sqrt{\sin^2\psi - n^{-2}}}{n^{-2}\cos\psi}\right) \quad (4)$$

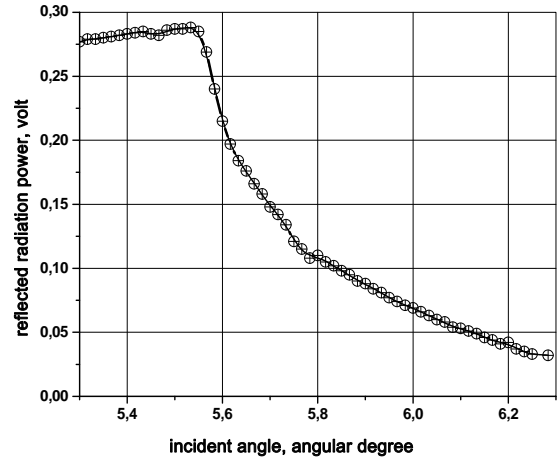

Fig. 3. The behavior of reflected beam power at the phase shift in the vicinity TIR on the 90°-degree prism. Air border TIR on a cathetus plane of a prism begins at incidence angle ≤5,55°

The manifestation of phase jump for p-polarized radiation in this example was registered on power change of the arising s-polarized components in a narrow sector of incidence angles for TIR appearance and normal reflection without phase change. Against photodetector, the analyzer to transmit arising s-polarized light component was installed. Multiple increases of the s-polarized part power of a reflected beam arising due to phase shift at TIR in narrow transitional sector 1° is provided on fig. 3. For RI =1,5 and ψ=45° estimated phase jump reaches ≅ 28,07°. (We do not provide data of similar calculation for s-components and their differences ($\delta_p - \delta_s$) in view of the small amplitude of the last).

Thus, arbitrary small power part of s-polarized radiation at reflection can always remains as due to of unattainability the absolute values of DP=100 % and α=0 for input light, and also owing to its depolarization, caused by light scattering on irremovable atomic and molecular scale roughness of the surface. Below we will show availability of very small light scattering at the reflection on the even atomic pure surface of mica plates. Light scattering from the liquid surface could make a contribution to the measurements taken by Rayleigh in the context of the analysis of this problem (without laser beams!).

In our example of fig. 2. level of the reflected s-polarization power on small incidence angles was below the power level of p-polarization in ≈30000 (more than 4 orders). S-polarized parts of the initial radiation could arise as from limited values of DP of the laser source and azimuthal orientation its linear polarization (1-p≠0, α≠0), so as light scattering on multiple optical surfaces of setup. The addition of the radiation power presented by curves 1P and 2S of fig. 2. leads to small to shift (units of minutes) of MRP for Brewster angle and, respectively, to an error of determination of RI using Fresnel's formulas. Therefore, application of the best analyzers (with the relation of transmissions of orthogonal components ≥1000) for suppression s-polarized components of the reflected light is a necessary condition of receiving correct measurements of Brewster angle with MRP angular position. The power level of MRP in fig. 1. without polarizing filtering of s-component in the case was equal ≈33mv. With the installation of an analyzer power level in MRP fell to ≈4mv, its angular position was shifted to 56,545º that improved the measured RI value to 1,5134 with an agreement to state standard specification in the 4th sign.

### 3. Determination of Brewster angles of thin-layer materials

MRP generation in the reflected light for a p-polarized incident beam in the case of thin plates and films is accompanied by an interference of two beams of almost equal intensity. Though reflection by a back surface can be suppressed with an antireflection coating, roughening or blackening of back surface to attenuate the second beam, this case is necessary to consider in more detail because of a whole set of the various possibilities.

Thin layer materials take huge family film layers on substrates, various free films and thin plates made of absorptive and transparent materials. Specifics of MRP finding for it and its identification with Brewster angle in the presence of 2 beams reflected from the plane-parallel front and back surfaces appear in the general case as two-beam interference in parallel beams. If testing wavelength suffers from the noticeable light absorption or scattering in at a volume of film or on its back surface, the second beam is suppressed and a problem of determination of RI can be reduced to described above procedure. In the presence of wedge angle between surfaces reflected beams leave under the angles differing at a wedge angle and therefore physically equal angles of MRP of this film will be displaced on a size of this angle magnitude. No interference in parallel beams for the case even at small wedge ~ 1' can be registered.

The similar results for the wedge plate from K8 glass is shown in fig. 4. The interference of equal thickness for a wedge plate is in the plane of the film, but at separate beam development it does not influence the MRP provision on each of beams. Registration of total radiation of two beams will bring in a similar case to mixed MRD from two Brewster minimum. However for a wedge plate it always possible to find the beam reflected by the front plane and definitions of true Brewster angle if a wedge magnitude is unknown.With the reduction of film thickness to several microns parallelism of the planes become better, and the interference in the parallel beam (at an equal angle of incline) shows the higher visibility because it is strongly sensitive on wedge angle value [11].

In general the two-beam interference in the reflected light from a plane-parallel plate is characterized by the space period, depending on wavelength λ, plate thickness T and RI, incident/refraction angles φ/ψ [7]. The angular period as a difference of refraction angles for the alternating maxima is defined by dependence (5).

$$(\cos\psi_1 - \cos\psi_2) = -2\sin\psi_{ave}\sin\delta\psi = -\frac{2}{n^2}\sin\varphi_{ave}\sin\delta\varphi = \frac{\lambda}{2T}$$

$$\delta\varphi \approx \frac{\lambda n}{4T\sin\varphi_{ave}}$$

(5a,b)

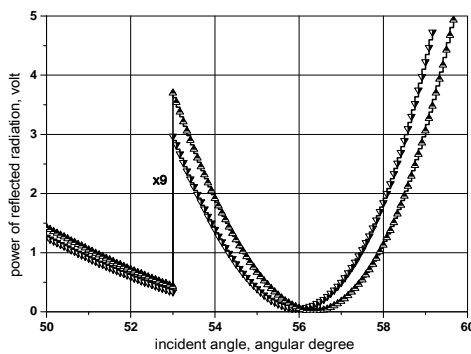

Fig. 4. MRP positions for a wedge plate with angle 25' (jump of power on 53º is caused by 9-fold increase of input power to improve signal/noise relation in vicinity of MRP (K8 glass))

The formula (5a) represents a difference of angular positions of interferential maxima through the angles of refraction/incidence $\psi/\varphi$ and an optical thickness of a plate of nT. The approximate formula of the angular size of the space period is given by a formula (5b). The space period depends from angular period and distance between the tested plate and a photodetector. It is clear that film thickness T, RI≡n and angles $\psi/\varphi$ determine the angular period of an interference and its relation to the tested beam divergence. Other details of the interference manifestation in parallel reflected beams have been considered in our work in connection the special coherence measurement [10]. In the context of this study – influence of interference on RI measurement by Brewster refractometry - consider measurements of MRP and, respectively, RI in the presence of the mentioned two-beam interference for plates of fused silica and coverslip microscope glasses.

The space period of an interference on plates with T=1mm is equal ≈0,25' while the rotation step of our setup is 1 order smaller. Therefore if the plate is plane parallel is better than 1' interferential minimaxes should be observed and registered, however, that does not happen (see fig.5.) Registered MRP corresponds to fused quartz: $\varphi_{brw}$ =55,5⁰ and RI(660nm)=1,455.

As thickness reduces to 150mkm (wedge decreases) interferential pattern appears and is already registered: its amplitude and the angular period decrease with rotation to bigger incidence angle in accordance expression (5), at last disappears quite in the vicinity of Brewster angle (fig. 6.), without any obstacles for its correct determination. On insert to fig. 6 increase of the registered visibility of the interference of this sample at some increase of spatial resolution of a photodetector with the application of an entrance slit of 0,5 mm is shown. It is seen that owing to the lowest quality of optical surfaces of the coverslip and, respectively, stronger light scattering, the power of s-polarized components on the similar plates increases. However, its cut-off with the polarizing filter improves MRP angular position indirection to real Brewster angle for RI of K8 glass: $\varphi_{brw}$ =56,39⁰ and RI=1,504.

Thus, the main conclusion which follows from last results at manifestation of a two-beam interference and relative small light scattering on both sample surfaces consists in the feasibility of the successful Brewster refractometry of similar films and plates.

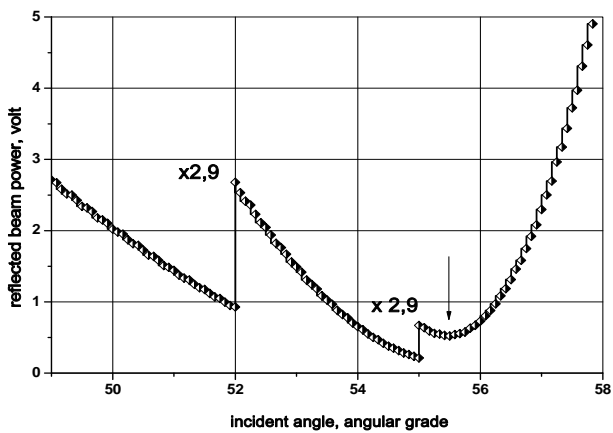
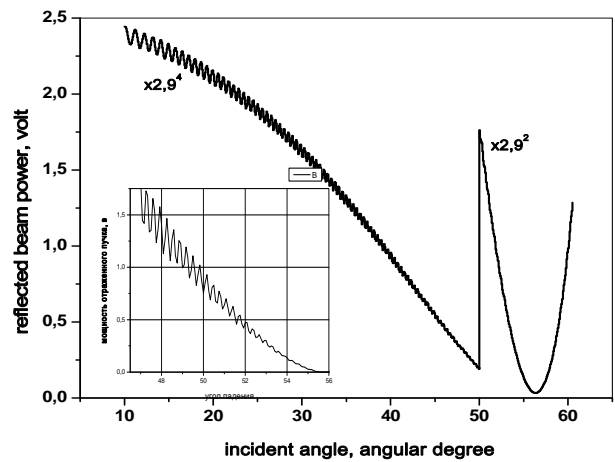

Fig.5. MRP for the quartz plate T=1050mk      Fig.6. MRP for glass plate T=150mk
                                                under interference manifestation

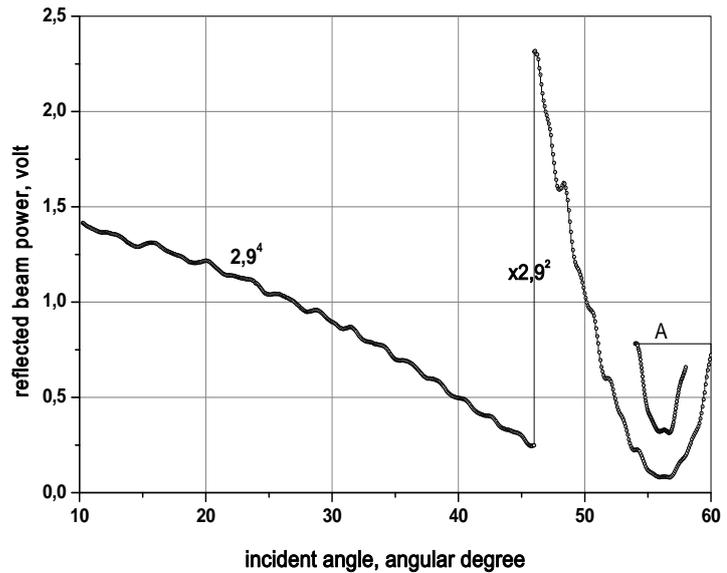

Fig.7. MRP for polyethylene film (T=32mkm) on glass substrate measured with diode laser λ=660nm. Value $2{,}9^4$ indicates the attenuation of the laser beam.

Similar successful RI measurements allowed to apply Brewster refractometry to films without any optical preparation of surfaces at all and even absolutely nontransparent due to light scattering material like milk or fluoro plastic. The similar nontransparent materials due to internal light scattering remove the two-beam interferometry totally. Instead, they need the single small part of the surface to be optically polished for Brewster angle measurement. Determination of MRP and RI of a polyethylene film with T= 32mkm provided the following data: φ=56,333° and PP=1,501 that is typical for the low pressured polyethylene (fig.7.). The curve A on the fig.7. corresponds MRP measurement at the increased power of the testing beam. Small waviness on the curve arises to residual two-beam interference. Its low-visibility resulted from above mentioned poorest boundary conditions for interference, in particular, caused also by deceleration of power of the second reflected the beam from a back surface of a film owing to immersion by a glue layer of the substrate.

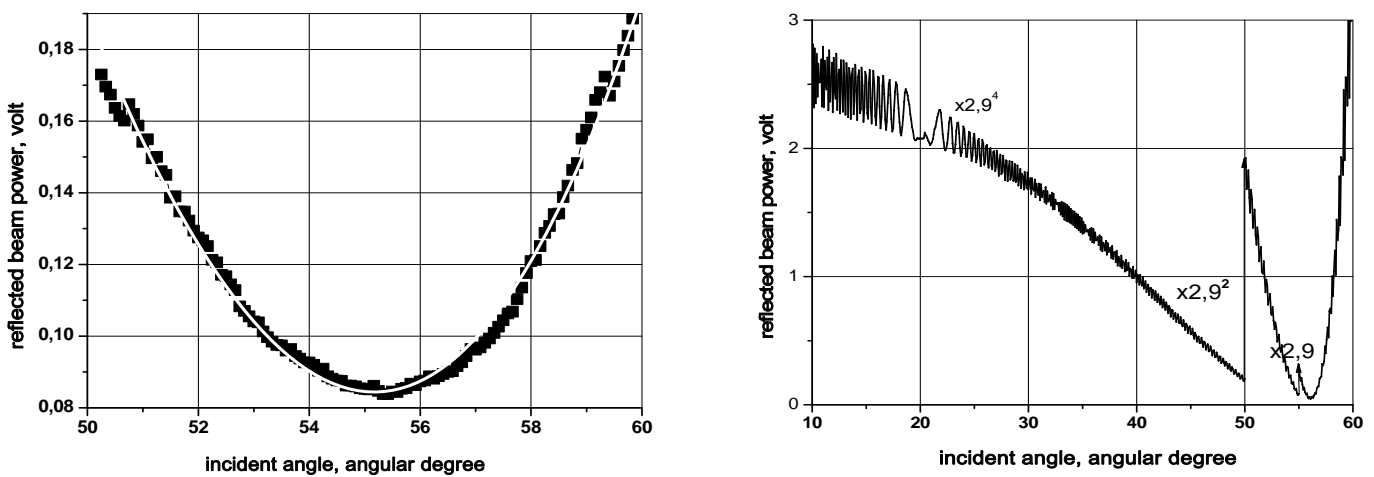

Fig.8a(left). MRP of dyed structured silicate film in absorption band R6G on 532 nm. The white line the results from regressive processing of experimental data with the 3rd power polynomial: φ$_{mrp}$ =55,20° and n=1,438; fig.8b(right)- MRP and Brewster angle of the same sample on 660 nm outside absorption band are φ=56,15° and n=1,491.

Brewster measurement RI for samples of submicron thickness (including absorbing) presents interesting in general and particularly for the silicate structurally arranged films prepared with sol-gel technology with the use of polymeric surfactant. The arranged pores of such films filled with polymer and were painted by R6G dye. The thickness of films measured by the method of atomic field microscopy was determined in limits (180-200)nm [8]. RI Brewster measurements were taken on wavelengths 532nm and 660nm in order the smaller one to match an absorption maximum of R6G while bigger one lay outside its absorption band.

At very high molar concentration R6G in a film sample of nanosized thickness the absorption coefficient became so high [8] that the real and imaginary part of a complex RI $\varkappa$ became commensurable. Measurements of complex RI with Brewster angle are unambiguous, in opposite case when $n >> \varkappa = \lambda\alpha/2\pi$. In fig.8a, results of measurement of angular position MRP for the specified samples are given. The sample absorption actual for MRP measurements on $\lambda$=532nm removed the second beam and an interference; measurement of MRP outside the absorption band of dye matched real Brewster angle of silicate matrix and was followed by an two-beam interference on $\lambda$=660nm (fig.8b): the main and reflected from a back surface of a substrate on which the "nanometric" layer was grown up (the back surface of this layer is antireflection by quartz substrate).

The angular standing of MRP for wavelength 532nm under total contribution of n and $\varkappa$ is described with exact but bulky Fresnel expression. If one use its approximate solution [9], the real part RI can be determined by the angle of MRP and independently from measured absorption coefficient of α. Really, when inequality of $A=n^2(1+\varkappa^2)>1$ [9] and $B=2n$ perform an angular dependence of a power reflectivity for the p-polarized radiation (TE-wave) can be provided by the following expression:

$$R(\varphi) = \frac{A\cos^2(\varphi) - B\cos(\varphi) + 1}{A\cos^2(\varphi) + B\cos(\varphi) + 1} \qquad (6a)$$

The derivative extremum on the angle j of function (6a) allows to find angular position MRP:

$$\varphi_{brw} = \arccos(1/\sqrt{A}) = \arccos(1/n\sqrt{(1+\varkappa^2)}) \qquad (6б)$$

Data for MRP angle on fig. 8a permit to find RI≡n=1/cos($\varphi_{brw}$) $\sqrt{}$(1+$\varkappa^2$)=1,437 при $\varkappa$=0,7.

For next understanding, the origin of the nonzero reflected power on Brewster angle on TE incident radiation material with the atomic pure surface (no light scattering) presents interesting. It is known that surface of mica plane is natural cleavage of a crystal structure. We used muscovite plates with a thickness of 80mk. Muscovite belongs to a monocline syngony and is optically two-axial material. The plane of the optical axis and the plane of cleavage are orthogonal with an angle between optical axis ≈40°. RI along optical axis differ slightly: $n_g$=1,613 ÷ 1,596, $n_m$ = 1,607 ÷ 1,596, $n_p$=1,569 ÷ 1,561 and, respectively, is small birefringence 0,038 ÷ 0,045 [10]. Values of Brewster angle for optically of anisotropic material depends nontrivially on mutual orientation of an optical axis crystal and the incidence angle of the testing beam. The scheme of registration with the analyzer on the photo receiver, selecting p-polarized emission (TM), gave up hope to find MRP connected with an ordinary beam.

Results of measurement are presented on fig.9. The MRP angle occurs to be 58,1660 for λ=660nm corresponds to RI value equals n=1,610 which is inside of limits to $n_g$=1,596-1,613.

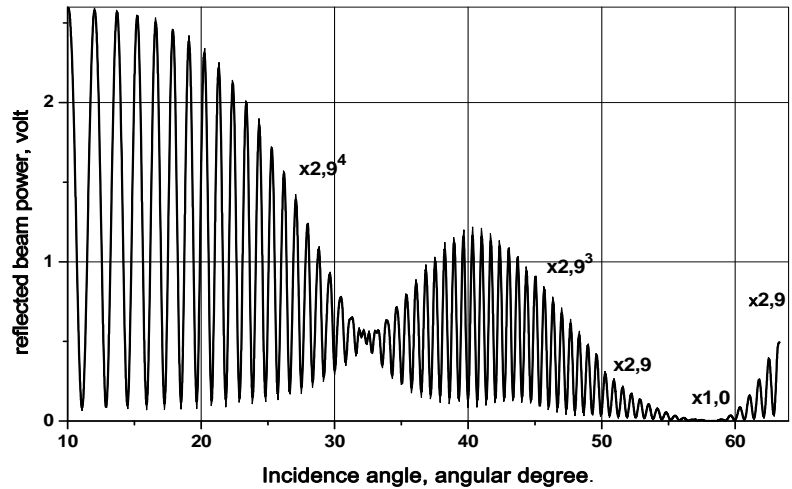

Fig. 9. Angular dependence of the reflected power of the p-polarized wave 660 nm for the atomic pure surface of muscovite, T= 80mk. On Brewster angle, the power of incident radiation is increased by $2,9^5$ times in order improve signal/noise relation in MRP vicinity.

Dependence in fig. 9. contains the following various information:

a) The interference with the contrast of visibility close to 100% is imposed on the angular dependence of reflected light radiation which is linearly polarized p-configurations (TE) and specifies the availability of Brewster angle.

c) The angular period of an interference is described by the stated above ratio (5b) brought out of interference model in parallel beams for equal slope angles: change of the angular period is obliged to change of an incidence angle of the paraxial light beam on a plate.

c) The high contrast of interference visibility is connected only with the high optical quality of surfaces of a plate and its power of plane-parallelism i.e. good quality inherent to an interferometer [12].

e) This rotary interferometer is capable of measuring the width of space coherence of a measuring beam owing to imposing of the copies of a measuring beam arising in reflection and which are moving apart with the growth of a turning angle of a plate [12]. Therefore, the contrast of visibility with the growth of an incident angle typically decreases.

The first minimum on fig.9. on then incidence angle 32,26° is not connected with above mentioned effect of spatial coherence of used diode laser 660nm and can be related to accidental implementation of phase shifts between interfering beams with formation of s–polarized radiation component which is cut off by the analyzer (at turn of the analyzer on 90° it is confirmed by emergence of a power maximum on this angle). The registered MRP parameters for the mica strikingly differ from similar one for plates of even optical quality (fig. 4,5,6) in the extremely high visibility of an interference that is connected with the lowest level of light scattering and on its planes of cleavage and due to a very small angle wedge of the plate. The s-component power of the reflected radiation in the narrow vicinity of Brewster angle decreased a noise level even at the maximum power of the applied source of the p-polarized radiation (more than 10 mW). Distinction in the power of the reflected radiation on angles 10º and Brewster is not less than 3 orders!

## 3. Final provisions and conclusions

The conducted research contains the two connected parts: metrological part of the content is a study of determination opportunities of RI different materials and its topology with Brewster angle refractometry and the fundamental part connected to study of the residual radiation origin at light reflection at Brewster angle. It is shown that highest requirements to measuring light on the limit extent of linear polarization and installation of a zero azimuthal angle for a precision finding of Brewster angle are practically implemented at the installation in front of light receiver the p-polarization analyzer (the filter with a cutoff of s-polarized components in a reflected beam). Comparison of the reflected radiation integral on polarization to differential one indicates following: the nonzero azimuthal angle and the final degree of polarization influences on the angular position MRP on the same. Similar comparisons of the reflected power of the incident p-polarized radiation for s,- and p-polarized components indicates the source of an origin of s-polarized component registered in the vicinity of Brewster angle: this source is caused by optical heterogeneity of a surface unremovable by optical polishing and even when using an atomic pure crystal surface. Polarization degree change of the reflected radiation in the vicinity of Brewster angle is not accompanied by phase jump on $\pi$ radian what testifies the monotonous change of residual radiation power. Experimental observation the similar phase jump at TIR is followed by growth of power in a transitional area that confirms a correctness of the used measurement technique and conclusions concerning the nature of residual radiation on Brewster angle. Thus, the hypothesis of an intermediate super thin layer and the additions made Drude's model to basic Fresnel theory for explanations of the residual radiation origin lose the relevance.

The RI measurements executed on thin plates and films even with imperfect surfaces and anisotropy showed that an interference under these conditions does not prevent a finding of Brewster minimum because in the respective angular area an interference visibility decreases to zero. Light scattering level on imperfect surfaces, when the mirror component of light scattering after reflection still remains, does not interfere to determination RI on MRP due to the registration scheme with the polarizing light filter.

The possibility of RI measurement of thin and over thin films under T≤200nmм, including a complex RI is also possible: the registered MRP determines the real part of RI, but independent measurement of an absorption (imagine part RI) is necessary.

Unexpectedly successful were measurements RI of anisotropic films and plates with technological (tension) and natural birefringent like to polyethylene and mica. In particular, the defined values RI needs some theoretical justification of its binding to one of 2 optical axes of 2-axis crystal. The appeal to mica has evoked a possibility of comparative assessment of very low light scattering on the atomic pure surface of cleavage and light scattering on surfaces with optical polishing (or without that) on the determination of true Brewster angle from MRP. Extremely high visibility of the interference pattern in the reflected light at an extremely low power of s-polarized component on Brewster angle in comparison with all other samples indicates level the importance of light scattering in such measurements.

The following statements are the main conclusions of all presented work:

- The analysis of angular dependences of reflected beam power integral and differential on polarization the s, p-component indicates a lack of the polarization change predicted by model Drude p-s-p transition in the small vicinity of Brewster angle. It allows to offer an alternative explanation of the observed change of polarization in the specified sector in the expense of a contribution of a nonvanishing component of the s-polarization arising in a reflected beam.

- Closest to Brewster angle values of angular MRP position, at all possible decrease in the extent of linear polarization of the testing radiation and deviations of its azimuthal angle from zero value, are reached with the installation of the polarizing filter (analyzer) providing the strongest suppression of s-components in the reflected radiation.

## 4.The quoted literature

1) E.A.Tikhonov, A.K.Lyamets, Proceeding of Advanced Optoelectronics and Lasers (CAOL), 2013 International Conference, 2013.
2) E.A. Tikhonov, V.A. Ivashkin, A.K. Lyamets, Journal of Applied Spectroscopy 03/2012; 79(1). DOI:10.1007/s10812-012-9577-3
3) Lj.Arsov, M.Ramasubramanian, B.N. Popov, Ellipsometry, (2003), DOI: 10.1002/0471266965.com 062.pub2, Copyright by John Wiley & Sons, (2012)
4) D.V. Sivukhin., General course of physics, Moscow, Science, (1980), 751 pp.
5) P. Gyoll, How to turn the personal computer into a measuring complex, Publishing house of DMK, Moscow, (2005), 133pp.
6) I. L. Fabelinsky, Molecular scattering of light, Moscow, Nauka Publishing house, (1965), 400 pp.
7) Robert D. Guenther, Modern Optics, John Wiley and Sons, USA, (1990), 535 pp.
8) E.A. Tikhonov, G.M. Telbiz, Liquid Cryst. Mol. Cryst., (2011), v.535, p.82-92.
9) R. Dichbern, Physical optics, Moscow, Publishing house "Science", MRPML , (1965), 637pp.
10) Mica, types and key parameters, ГОСТ 10698-80, (1981)
11) A. Jain, S. Rogojevic, S. Ponoth, N. Agarwal and others, Thin Solid Films, (2001),v. 513,p.398-399
12) E.A.Tikhonov, A.K.Lyamets, arXiv.org:1509.09191, (2015)